\documentclass[12pt,iop,revtex4,numberedappendix,appendixfloats]{emulateapj}

\usepackage{graphicx}
\usepackage{amsmath}

\newcommand{\etal}{\mbox{et~al.}}
\newcommand{\eg}{e.g.\@~}

\newcommand{\ie}{i.e.\@~}
\newcommand{\fig}{Fig.~}
\newcommand{\tbl}{Table~}

\newcommand{\lya}{Ly$\alpha$}

\begin{document}

  \title{A Correlation Between \lya~Spectral Line Profile and
    Rest-Frame UV Morphology}
  
  \author{Vivian U\altaffilmark{1,7},
Shoubaneh Hemmati\altaffilmark{1},
Behnam Darvish\altaffilmark{1}, 
Bahram Mobasher\altaffilmark{1},
Hooshang Nayyeri\altaffilmark{1,2}, 
Mark Dickinson\altaffilmark{3},
Daniel Stern\altaffilmark{4},
Hyron Spinrad\altaffilmark{5},
%Henry Ferguson\altaffilmark{5}, 
Ryan Mallery\altaffilmark{1,6} 
}
 
 \altaffiltext{1}{Department of Physics and
   Astronomy, University of California, Riverside, 900 University
   Avenue, Riverside, CA 92521, USA; \href{vivianu@ucr.edu}{vivianu@ucr.edu}}
 \altaffiltext{2}{Department of Physics and Astronomy, University of
   California, Irvine, CA 92697, USA} 
\altaffiltext{3}{National Optical Astronomy Observatory, Tucson, AZ
  85719, USA}
\altaffiltext{4}{Jet Propulsion Laboratory, California Institute of
  Technology, 4800 Oak Grove Drive, Mail Stop 169-221, Pasadena, CA
  91109, USA}
\altaffiltext{5}{Department of Astronomy, University of California,
  Berkeley, CA 94720, USA}
%\altaffiltext{6}{Space Telescope Science Institute, 3700 San Martin
%  Drive, Baltimore, MD 21218}
\altaffiltext{6}{Rebellion Photonics, 7547 South Freeway, Houston, TX
  77021, USA}
\altaffiltext{7}{UC Chancellor's Postdoctoral Fellow}

  \begin{abstract}
We explore the relationship between the spectral shape of the
\lya~emission and the UV morphology of the host galaxy
using a sample of 304 \lya-emitting $BVi$-dropouts at $3 < z < 7$ in the
GOODS and COSMOS fields. Using our extensive reservoir of high-quality
Keck DEIMOS spectra combined with $HST$ WFC3 data, we measure the
\lya~line asymmetries for individual galaxies and compare them to axial
ratios measured from observed $J$- and $H$-band (restframe UV)
images. We find that the \lya~skewness exhibits a large scatter at
small elongation ($a/b < 2$), and this scatter decreases as axial
ratio increases. Comparison of this trend to radiative transfer models
and various results from literature suggests that these high-redshift
\lya~emitters are not likely to be intrinsically round and symmetric disks, but
they probably host galactic outflows traced by \lya~emitting
clouds. The ionizing sources are centrally located, with the optical
depth a good indicator of the absorption and scattering events on the
escape path of \lya~photons from the source.  Our results find no
evidence for evolution in \lya~asymmetry or axial ratio with look-back time.
  \end{abstract}
  

  \section{Introduction}
  \label{Introduction}

Lyman $\alpha$ emission (\lya~$\lambda$1216), due to the electron
transition from the second to the ground level of the Hydrogen atom,
is often the primary feature available to confirm high-redshift
galaxies and to probe their nature. Because of its short wavelength,
\lya~is conveniently shifted to optical or near-infrared wavebands at
high redshifts and hence, is accessible from ground-based telescopes.

The spectral profile of \lya~is indicative of the origin and escape
path of \lya~photons, as well as the resonance scattering and
extinction by neutral hydrogen and dust in the interstellar medium of
galaxies. Specifically, the \lya~line is often observed to be
asymmetric~\cite[][]{Stern99,Verhamme06},
double-peaked~\cite[][]{Ahn01,Tapken07,Fosbury03,Christensen04,Venemans05,Kulas12,Yamada12,Laursen11}, 
and/or exhibit a P-Cygni profile~\cite[][]{Dickinson98,Dey98,Bunker00,Ellis01,Ahn03}.
The different physical origins of \lya~photons from high-redshift
star-forming galaxies include hot, ionizing sources (\eg AGN or massive OB
stars in young star clusters or ionized circumstellar regions),
cooling from gas heated by shocks from gravitational collapse, or by
shocks due to outflowing gas from starburst regions or
AGNs~\cite[][]{Yamada12}. The temperature, density, and motion of the
surrounding medium may therefore affect the profile shape of the
\lya~emission~\cite[][]{Ahn01,Tapken07}. Additionally, processes
external to the galaxy, \ie absorption from the intergalactic medium,
may also introduce asymmetry to the line as \lya~forest opacity
increases at higher redshifts~\cite[\eg][]{Rauch97}. The real situation is likely
significantly more complicated, with \lya~emission also depending
on the properties of its host galaxy and in particular on its
interstellar environment.

To date, several large, dedicated spectroscopic surveys of
high-redshift \lya~emitting sources have investigated the profile
shape of \lya. Many studies have included quantitative comparison of the
\lya~line profile morphologies to predictions from various radiative
transfer
models~\cite[e.g.][]{Verhamme06,Verhamme08,Dijkstra07,Laursen11,Yamada12}. For
instance, \cite{Ouchi10} studied a sample
of 207 $z$ = 6.6 \lya-emitters (LAEs) in the Subaru/\emph{XMM-Newton} Deep
Survey field and found no large evolution of the \lya~line profile
from $z = 5.7$ to 6.6. A similar finding was confirmed by~\cite{Hu10}
using a sample of 118 $z \sim 5.7$ and 6.5 LAEs selected from a
narrowband survey of the GOODS-North field. 
\cite{Mallery12} investigated the \lya~equivalent widths and escape
fractions of 244 sources with \lya~emission at $z \sim 4-6$ and found
no redshift dependence; instead, dust extinction is the major factor
that inhibits the escape of \lya~photons.  These large 
samples provide a basis for understanding the observed
spectroscopic properties of high-redshift LAEs. 

In order to further investigate the escape path of \lya~photons
from high-redshift LAEs, we explore the link between their
spectroscopic and morphological properties, partially motivated
by a similar relation studied by~\cite{Law12}. Specifically, we examine the
relation between \lya~spectral profile shapes and rest-frame UV
morphologies for a large sample of 304 LAEs (spectroscopically
identified by their \lya~line) at $3 < z < 7$ as well as
the evolution of the line profile shapes with redshift. By examining
these independent representations of the UV emission, we expect to
gain insight into the internal and external processes as well as
geometrical and projection effects that may
govern the observed shape of \lya~line profiles.  We do so by
using a spectroscopically selected sample of high-redshift
\lya~emitting galaxies in the Great Observatories Origins Deep
Survey~\cite[GOODS;][]{Giavalisco04} and the Cosmic Evolution
Survey~\cite[COSMOS;][]{Scoville07} fields observed with
DEIMOS~\cite[][]{Faber03} on the Keck II Telescope. The available
$HST$ ACS and WFC3 images along with multi-band photometric catalogs
from COSMOS and the Cosmic Assembly Near-infrared Deep Extragalactic Legacy
Survey~\cite[CANDELS;][]{Grogin11,Koekemoer11} provide valuable
information regarding the morphology of the rest-frame UV emission. 
  
  This paper is structured as follows. The
  sample selection is described in \S \ref{Sample}. Section
  \ref{Analysis} details the \lya~line fitting and 
  morphological analysis of galaxies, while \S \ref{Results}
  presents the line profile asymmetry and elongation measurements.  The implications
  of our results are given in \S \ref{Discussion} and summarized in \S
  \ref{Summary}. We assume $H_0 = 70$\,km\,s$^{-1}$\,Mpc$^{-1}$,
  $\Omega_{\rm m}$ = 0.3, and $\Omega_\Lambda$ = 0.7 throughout the paper.

  \section{Spectroscopic Sample Selection}
  \label{Sample}
  The high spectral resolution of the present spectroscopic data set
  makes it well suited for a study of the \lya~line profile shape. For
  both GOODS and COSMOS, the high-redshift LAEs have been observed
  with the optical multi-object spectrograph DEIMOS
  on the Keck II Telescope.  The DEIMOS observations were taken using
  1\arcsec~slits with either the 600 $l$/mm or the 830 $l$/mm gratings
  (full width half maximum resolution of 3.5\AA~and 2.5\AA,
  respectively) spanning a wavelength range $\sim 5,000-10,000$ \AA,
  designed to detect \lya~lines at high redshifts. Given the redshift
  range of our sample ($3.22 < z < 7.21$), the spectral
  resolution at \lya~is $R = 1470-4000$, corresponding to velocity
  resolution $\Delta V = 75-200$ km s$^{-1}$. Integration times per
  mask varied, but was typically on the order of two hours per mask.

\subsection{The GOODS Sample}
  The Keck spectroscopic follow-up campaign of high-redshift galaxies
  (largely selected as Lyman break galaxies, or LBGs, as well as from
  photometric redshifts and narrowband surveys) in the
  GOODS+CANDELS fields has been an on-going project since 2005.
We used the DEEP2 pipeline to process the DEIMOS data. The observed 2D
spectra were sky line subtracted and flat fielded. The observed
standard lamps were used to construct a wavelength solution and to
wavelength calibrate the spectra. The individual 1D spectra were
extracted from the reduced 2D using nominal positions from the mask
design files using boxcar and optimal extractions. For the 830 $l$/mm
grating, observations were dithered to correct for the ghost
effect and for better skyline subtraction. In this case, a modified
version of the pipeline was used 
where the data were processed using the dithering
pattern recovered by the pipeline from observations of the bright
stars on the mask. The final output are 1D and 2D spectra for
individual sources, wavelength calibrated and skyline subtracted.

  All the reduced spectra in the GOODS sample have been visually
  inspected by at least two experienced team members for redshift
  identification. From the pool of sources at the appropriate redshift
  range where \lya~is observed, we applied a detection 
  limit of $3\sigma$ in \lya~line flux which results in the selection of 130
  LAEs with secure (quality code $Q = B$; well-identified \lya~line)
  and very secure ($Q = A$; well-identified \lya~and additional
  emission or absorption lines) redshifts. Ninety-one of the LAEs are
  in the GOODS-North field (with only five sources matched to
    AGN identified in a 2-Megasecond X-ray catalog; \emph{private
    communication with D. Kocevski}), while 39 are in the GOODS-South
  field. These sources were matched to the CANDELS photometric catalog
  and encompass the sample of LAEs from GOODS considered in this study.   

  The LAEs were further categorized by their Lyman break properties.
  Previous photometrically selected samples have allowed us to probe the
  UV luminosity function at $4 < z < 8$~\cite[\eg][]{Bouwens10} and
  raise questions about the reionization state of the intergalactic
  medium (IGM). Since the \lya~emission fraction in LBGs depends on
  the IGM neutral fraction~\cite[][]{Stark11}, it is crucial to clarify the spectral
  properties of \lya~in LBGs as identified by dropout
  techniques. Dropout techniques utilize images taken in multiple
  photometric wavebands to detect the Lyman 
  break in high-redshift sources~\cite[][]{Steidel96,Dickinson98}. For
  instance, the following color-color selection as adopted from, \eg
  \cite{Beckwith06} and \cite{Stark09}, are ideal for identifying
  candidate star-forming galaxies at $z > 3$: \\
  \\
  $B$-dropouts: 
  \begin{multline}
    B_{435}-V_{606} > 1.1 + V_{606} - z_{850}, \\
    B_{435}-V_{606} > 1.1,  \\
    V_{606} - z_{850} < 1.6, \\
    S/N(V_{606}) > 5, \\ 
    {\rm and}  S/N(i_{775}) > 3.
  \end{multline}
  $V$-dropouts: 
  \begin{multline}
    V_{606}-i_{775} > 1.47 + 0.89 (i_{775} - z_{850}) \quad {\rm or} \quad 2, \\
    V_{606}-i_{775} > 1.2, \\
    i_{775}-z_{850} < 1.3, \\
    S/N(z_{850}) > 5, \\ 
    {\rm and} \qquad   S/N(B_{435}) < 3, \quad {\rm or} \quad B_{435}-i_{775} > V_{606}-i_{775}+1.
  \end{multline}
  $i$-dropouts: 
  \begin{multline}
     i_{775}-z_{850} > 1.3, \\
    S/N(z_{850}) > 5, \\ 
    {\rm and} \quad S/N(V_{606}) < 2, \quad {\rm or} \quad V_{606}-z_{850} > 2.8.
  \end{multline}

\begin{figure*}[hbt]
  \vspace{-2.4in}
  \centering
  \includegraphics[width=.7\textwidth,angle=90]{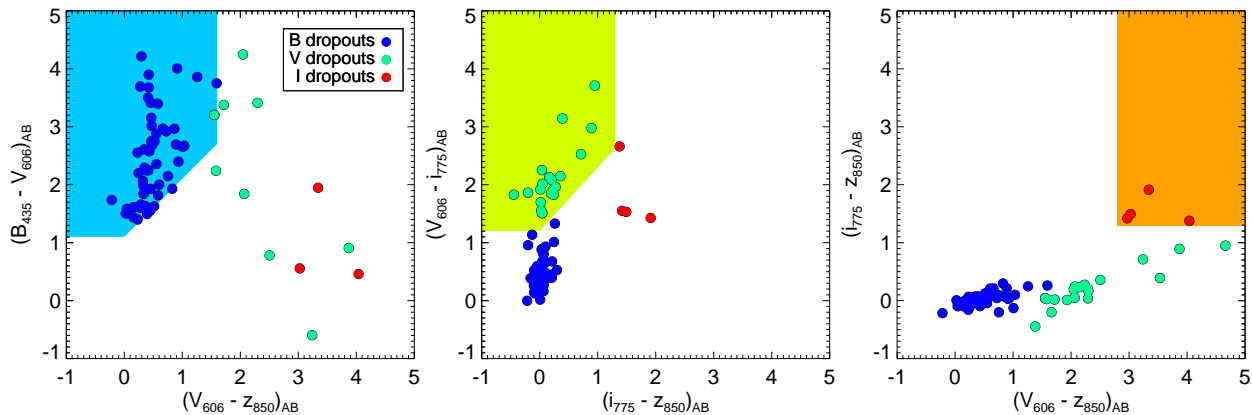}
  \vspace{.2in}
  \caption{Color-color diagrams for the LAE sample in the GOODS
    fields, showing LBG color selection of $BVi$ dropouts as indicated
    by the shaded areas in each panel. Of the 130 sources identified
    with the \lya~line (black circles) in GOODS, 61 are categorized as  
    $B$-dropouts (blue circles), 21 as $V$-dropouts (green circles; one of which
    is also a $B$-dropout), and eight as $i$-dropouts (red circles).}
  \label{fig:ccd}
\end{figure*}

  Using the available photometry catalogs for GOODS~\cite[][Barro
  \etal~in prep.]{Guo13}, we categorize the LAEs by their Lyman
  break properties.
  The LBG color selection thus results in 57 $B$-dropouts, 20 $V$-dropouts (one of
  which is also a $B$-dropout), and 6 $i$-dropouts. The
    remaining 47 sources fail to qualify as a dropout due to one of
    two reasons: either photometric errors in their broad-band
    photometry scatter them in color-color space, or that the
    \lya~equivalent width is so large it affects their broad-band
    colors used to determine their LBG status. The LAEs are shown
  on the color-color diagrams in \fig 
  \ref{fig:ccd}. 

  \subsection{The COSMOS Sample}
  The COSMOS sample of LAEs were initially presented
  in~\cite{Mallery12}, with similar spectroscopic observational setups
%  (1\arcsec~slits with the 830 line BK7 grating with wavelength
%  coverage of $\sim6000-9000$\AA)
  and reduction
  procedures as the GOODS sample. These sources were selected to have
  photometric redshifts $z > 3.8$ and were complete to $z$ magnitude
  $z^+ < 25$ and mass $> 10^{10.5} M_\odot$ (Mallery et al., in prep; Capak et al., in
  prep). Narrowband selections were also applied to identify
  LAEs~\cite[Scarlata et al., in prep;][]{Shioya09,Murayama07}. Of the 244 objects
  from~\cite{Mallery12}, 210 were matched to the
  available COSMOS morphology and photometry catalogs~\cite[][]{Leauthaud07,Capak07}. 
  Recent DEIMOS observations added 18 new LAEs, and
  altogether there were 174 sources that met the 3-$\sigma$
  \lya~detection criteria. Adopting the LBG selection in \S 2.2 
  of~\cite{Mallery12}, 9 were $B$-dropouts, 23 were $V$-dropouts, and one was an $i$-dropout.  

  \subsection{Properties of the Combined Sample}
\begin{figure}[htb]
  \centering
  \includegraphics[width=0.39\textwidth,angle=90]{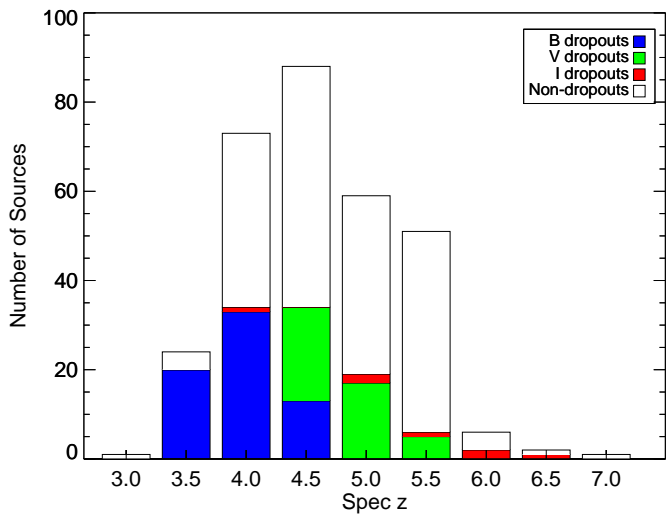}
  \caption{
%The redshift distribution of the total LAE sample (black),
%    also divided to different constituents: $B$-dropout (blue), $V$-dropout
%    (green), and $i$-dropout (red). 
    The stacked redshift distributions of the $B$-dropouts (blue), $V$-dropouts
    (green), $i$-dropouts (red), and non-dropouts (white) with bin
    sizes of 0.5. 
    Combined, the final sample spans the
    redshift range of $3.22 < z < 7.21$ with median $\langle z \rangle = 4.55$.}
  \label{fig:redshift}
\end{figure}

  Given the consistency between the observations and analyses of the
  GOODS and COSMOS LAEs, we combine the two populations for 
  a final sample of 304 LAEs  (including 66 $B$-dropouts, 43
  $V$-dropouts --- one of
  which is also a $B$-dropout, and 7 $i$-dropouts).  The redshift
  distribution is shown in \fig \ref{fig:redshift}, spanning the range
  $3.22 < z < 7.21$ with a median $\langle z \rangle =
  4.55$. Properties of the entire sample of LAEs can be  
  found in \tbl \ref{tbl:specprop}, while \tbl \ref{tbl:lbgsummary} 
  gives a summary of the dropouts.  We note that while the observational
  set up and analyses of the GOODS and COSMOS sources are consistent,
  the selections of the Lyman-break dropouts likely differ slightly
  for the two fields. The selection of LBGs in the GOODS fields was based
  primarily on existing multi-waveband images taken in the \emph{HST}
  ACS bands~\cite[\eg][]{Vanzella09}, whereas the selection in COSMOS
  relied on a combination of photometric filters used in the
  \emph{Subaru} Deep Survey~\cite[\eg][]{Ouchi04} and narrowband
  filters~\cite[see][for details]{Mallery12}. The depths in the
  equivalent bands between the two fields might also be different.
In this paper we do not directly compare the different
  selection criteria used for categorizing LBGs. We merely present
  their properties for the aggregate sample, and
  encourage curious readers to extract and consider the properties
  of those dropouts from their survey of choice. 

  We also point out that there are 190 sources
  classified as `Other' in our sample: these are \lya-emitting sources
  not classified as a dropout. The non-dropouts in GOODS are likely
  too faint for color selection, whereas those in COSMOS were
  primarily selected in narrowband surveys.  %At $z'$-band, the
%  nondropout sources are on average $\sim$0.14 mag fainter. 
  Thirty-nine of these 190 sources are not detected in
  $z'$-band, whereas all dropouts are detected.
  At $B$-band, the non-detection percentage among the nondropouts
  reaches 35\% ($106/304$), whereas that for the dropouts is only 18\%
  ($20/114$).  A detailed comparison between
  these galaxies and the LBGs will be presented in a forthcoming
  paper; we explore their similarities and differences briefly here.

  \section{Analysis}
  \label{Analysis}
  
  \subsection{Emission Line Fitting}
\begin{figure*}[tbh]
  \centering
  \includegraphics[width=0.9\textwidth]{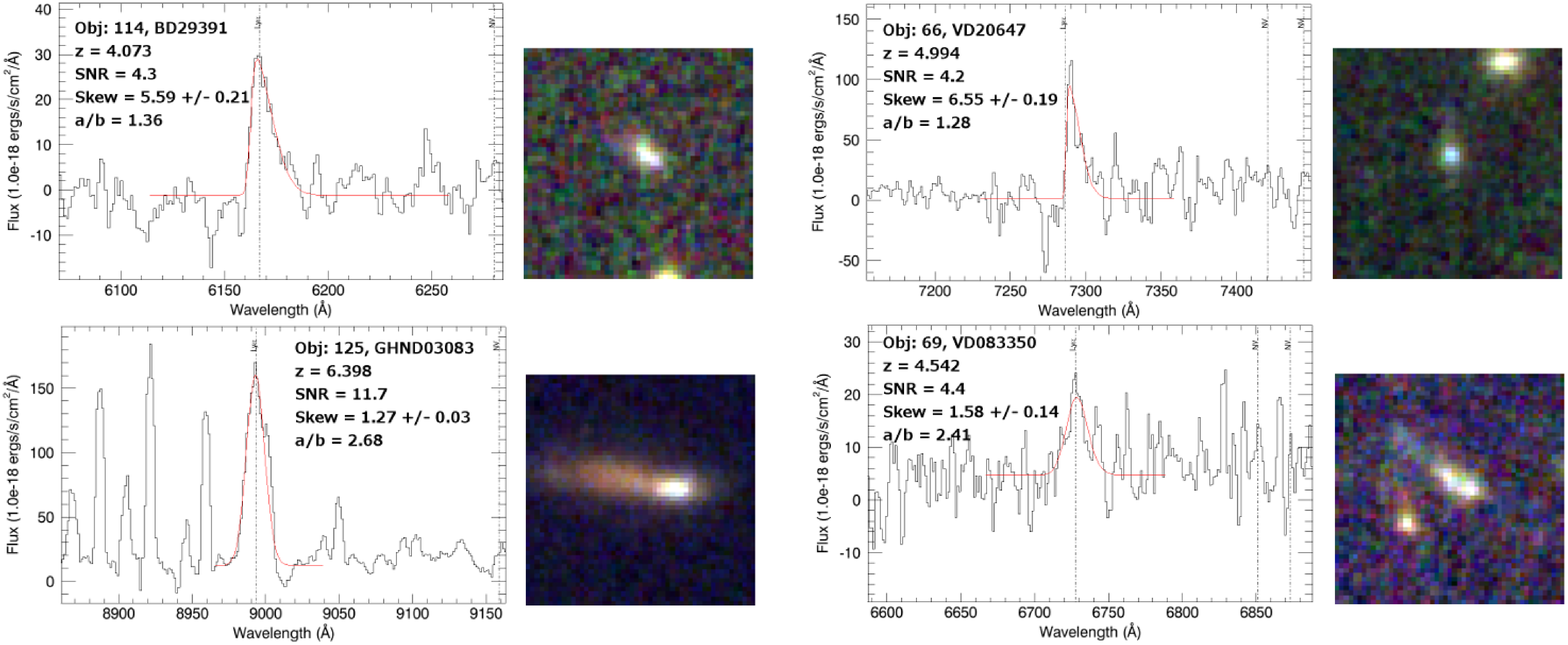}
  \caption{Examples of fitted \lya~lines along with their $zJH$-image
    in pseudo-RGB colors. The objects are selected to illustrate the
    span of skewness and axial ratio.}
  \label{fig:spim}
\end{figure*}

  All DEIMOS spectra were individually inspected following a
  detailed procedure outlined in~\cite{Mallery12}. As this is a
  spectroscopically selected sample, \lya~emission is seen in all the
  sources and has been used for redshift determination. We fitted the
  line profile of \lya~with a skewed Gaussian, and kept sources with
  at least a 3-$\sigma$ detection in \lya~flux over the noise in the
  continuum.  The functional form 
  of the skewed Gaussian was adopted from~\cite{Mallery12} and
  reproduced here: 
  \begin{equation}
    \text{flux} = A \times e^{-0.5\times((\lambda-x)/\omega)^2}\left(\int_{-\infty}^{s(\lambda-x)/\omega}\text{exp}(-t^2/2)dt\right)+c.
    \label{eqn:gauss}
  \end{equation}
  The fit returns values for the flux normalization ($A$), the first
  moment of a standard Gaussian ($\lambda_0 = x +
  \omega\delta\sqrt{2/\pi}$), the second moment of a standard Gaussian
  ($\sigma = \omega\sqrt{1-2\delta^2/\pi}$), the value of the skew ($s$),
  and the value of the continuum ($c$), where $\delta = s/\sqrt{1+s^2}$.
  For a symmetric line, skew $s$ would be 0. The sign of skewness
  demonstrates the directionality of the asymmetry, e.g. a large skew value
  indicates an asymmetric \lya~line with a large red wing, and a negative
  skewness denotes a blue wing. \fig \ref{fig:spim}
  illustrates some examples of the skewed and non-skewed \lya~lines
  at the lower signal-to-noise threshold (SNR $\lesssim$ 5) along with
  their corresponding fits. 
  An in-depth analysis to test the robustness of our
  measurements from line-fitting is presented in the Appendix.
  We report here the skewness and its associated uncertainty from
  the fit, as this quantity best describes the asymmetry of the
  \lya~line profile and could be used to trace the escape
  path of \lya~photons~\cite[][]{Yamada12}.

  \subsection{Morphological Analysis}
  To investigate the morphology of the galaxies from which \lya~photons
  escape, we examine the rest-frame UV emission using the $HST$ WFC3
  F125W and F160W ($J$ and $H$ bands, respectively, hereafter) images from
  CANDELS.  Given the redshift range of our sample (3.22 $< z <$
  7.21), we divided it into two bins ($z < 5$ and $z > 5$) such that
  the observed $J$- and $H$-band images correspond to the same
  rest-frame $\sim 2300$\AA~emission in the low and high redshift
  bins, respectively. This helps to minimize any effect in the geometry
  of the galaxy that might be due to the change in bandpass.  For
  galaxies above redshifts $z \gtrsim 5.5$, the choice of WFC3 filters
  over ACS ones is necessary to avoid having \lya~in the waveband,
  despite the better angular resolution due to smaller point spread
  function (PSF) full width half maximum and less undersampling
  afforded by ACS.  
  In addition, the $J$- and $H$-bands should not contain any
    significant line emission, and so they allow for pure measurements
  of the stellar continuum morphology. In contrast, the [OIII] line at
  5007 \AA~has been shown to be a common source of contamination in
  high-redshift LBGs and LAEs~\cite[][]{Stark13,Schenker13,Penin15},
  particularly for those sources with inferior quality flags.
  Therefore, we settled on using
  WFC3 images for our morphological analysis for consistency and
  reliability~\cite[\eg][]{Mosleh12}.
  We focus, in particular, on the axial ratio, or elongation, as a proxy
  for the morphology of the rest-frame UV emission. The shape of the
  galaxy, as described by the axial ratio, has implications for the
  viewing angle and, consequently, the amount of obscuration escaping
  \lya~photons encounter along our line of sight. 

  Using the WFC3 images from CANDELS, we ran Source
  Extractor over the near-infrared images to extract the axial ratios of the
  galaxies. Unfortunately, equivalent WFC3 images are not
  available for the COSMOS field outside of the CANDELS area where
  much of the current LAE sample reside. For the COSMOS fields, we have
  available $HST$ ACS F814W ($I$ band hereafter) images for all 
  LAE sources, and extracted the relevant mrophology
  from~\cite{Leauthaud07}. Thus, we calibrate $I$-band Source
  Extractor~\cite[][]{Bertin96} measurements by comparing the same
  $I$-band measurements to their $J$- and $H$-band counterparts in GOODS-North. 

\begin{figure*}[ht]
  \vspace{-1in}
  \centering
  \includegraphics[width=0.7\textwidth,angle=90]{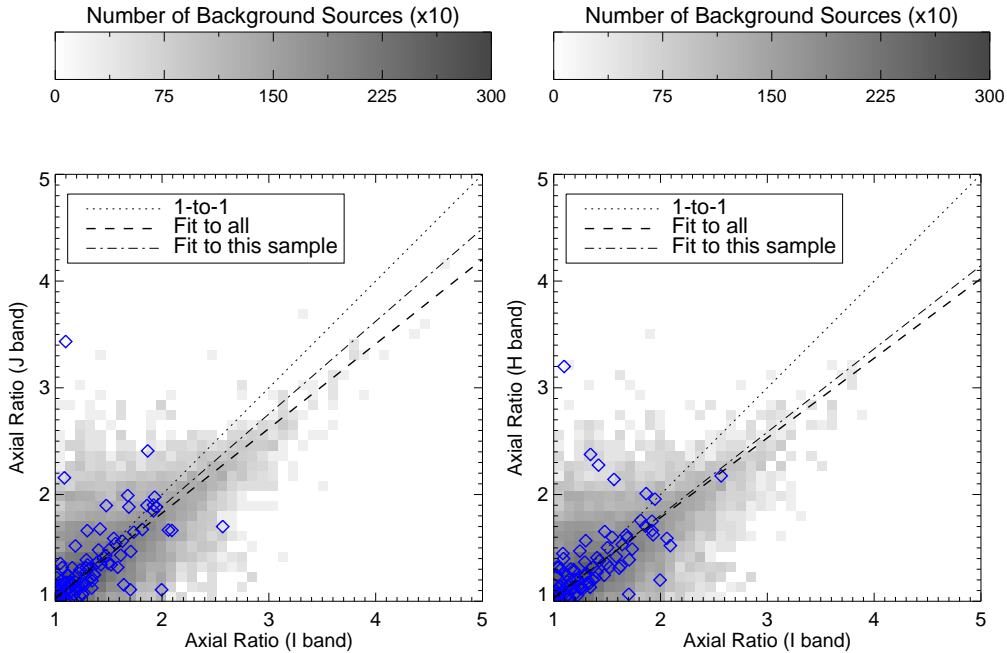}
  \caption{Comparison of the axial ratio measured in the $I$-band versus
    the $J$-band (left) and $H$-band (right) for all detected sources in
    the GOODS-N field. The blue open diamonds represent the LAEs in
    this sample.  The correlations are good between the bands, with
    negligible offset at the small end gradually increasing with
    elongation. Our sample (blue diamonds) exhibits smaller scatter
    ($\overline{\delta_{ij}} = 0.127$ and $\overline{\delta_{ih}} = 0.134$) 
    compared to the rest of the population ($\overline{\delta_{ij}} =
    0.133$ and $\overline{\delta_{ih}} = 0.142$).}
  \label{fig:elongIJH}
\end{figure*}

  Results in \fig \ref{fig:elongIJH}
  show that the offsets in $I$- vs. $J$-band and $I$- vs. $H$-band increase for
  larger elongations, and are almost negligible for smaller
  elongations. The mean scatter is smaller for our LAE sample
  ($\overline{\delta_{ij}} = 0.127$ and $\overline{\delta_{ih}} =
  0.134$)  than for the rest of the detected sources in GOODS-N
  ($\langle\delta_{ij}\rangle = 0.133$ and $\langle\delta_{ih}\rangle =
  0.142$). We note that the linear regression line for the detected
  sources in GOODS-N has a shallow slope ($< 1$), which is expected
  due to the PSF difference between WFC3 and ACS. The effect is
  pronounced at high axial ratios, where galaxies that are more
  elongated would look ``rounder'' in the WFC3 data due to the larger
  PSF. The underlying population also includes galaxies at lower
  redshifts (\eg $z < 2$) where one might expect rounder shapes if
  WFC3 samples the optical rest-frame compared to UV-rest frame for
  ACS, since rounder bulges become more prominent at redder
  wavelengths. This effect does not apply to the high-redshift LAEs in
  our sample (where only the aforementioned PSF effect would
  dominate), and thus the fit to our LAEs have a steeper slope
  compared to that for the underlying GOODS-N population, but still a
  shallow slope ($< 1$). Thus, the
  relations allow us to calibrate the $I$-band axial ratios ($a/b$) to
  that in $J$- and $H$-bands for the COSMOS sources as follows: 

  \begin{equation}
    a/b_{F125W} = 0.24 + a/b_{F814W} \times 0.79
  \end{equation}
  \begin{equation}
    a/b_{F160W} = 0.29 + a/b_{F814W} \times 0.75
  \end{equation}

  \section{Properties of High-Redshift \lya~Emitters}
  \label{Results}
  
    \subsection{\lya~Line Profile and Galaxy Morphology}
\begin{figure*}[ht]
  \centering
  \includegraphics[width=0.7\textwidth,angle=90]{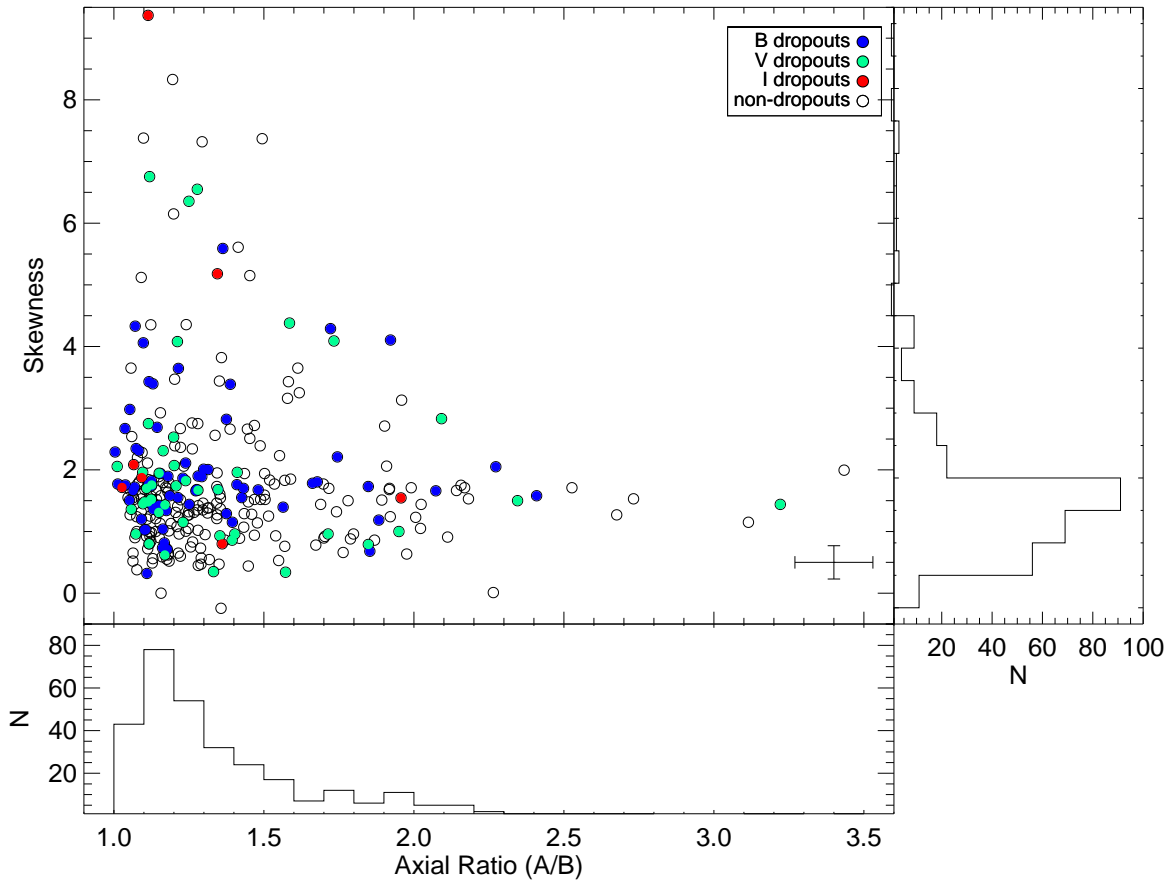}
  \caption{Skewness as a measure of the asymmetry of the \lya~emission
  line is plotted against axial ratio for all the LAEs
  (black). Subpopulations are color-coded by the
  dropout colors: $B$-dropouts (blue), $V$-dropouts (green), and $i$-dropouts
  (red). Typical error bars are shown in the lower right. The
  dispersion in skewness decreases from $\sigma = 1.44$ for $a/b <
  1.25$ to $\sigma = 1.27$ for $a/b > 1.26$ as elongation increases
  for the population.  The distribution in skewness and axial ratio are
  illustrated on the right and bottom, respectively.}   
  \label{fig:elongskew}
\end{figure*}

    We examine the relation between the \lya~spectral profile shape and the
    morphology of the rest-frame UV emission (as measured from the axial
    ratios) of the LAEs in \fig \ref{fig:elongskew}. For this
    analysis, we plotted all 304 LAEs in our sample, LBG or not, as black circles,
    and sub-populations of $B$-dropouts, $V$-dropouts, and $i$-dropouts are colored
    in blue, green, and red, respectively.  For low axial
    ratio values ($a/b < \langle a/b \rangle = 1.26$), a large dispersion in
    skewness ($\sigma = 1.44$) is found. As the elongation of the
    galaxies increases, this dispersion 
    decreases ($\sigma = 1.27$ for $a/b > 1.26$) and the skewness of
    \lya~converges to 1.53. More importantly than the measured
    dispersions, a large tail of galaxies with high 
    skewness is observed. No sources with large axial ratio exhibits large
    skewness in their \lya~spectral profile shape. The physical
    interpretation of this result is discussed in \S \ref{Discussion}.  

    The skewness distribution of LAEs is shown on the right panel
    of \fig \ref{fig:elongskew}. %, also sub-categorized for the $BVi$ dropouts. 
    The median skewness of the dropouts is 0.03-0.26 elevated
    compared to the entire sample, but otherwise the
    distributions are similar (see \tbl \ref{tbl:lbgsummary}). We
    employ the Kolmogorov-Smirnov (KS) test to determine the
    probability that the dropout and non-dropout populations are drawn
    from a similar distribution. Results from the KS tests among the various
    dropout groups return $p$-values of $0.15-0.97$, so any difference
    between these distributions is far from significant. 

    We consider how our sample's skewness distribution compares with
    that from~\cite{Yamada12}, who presented the \lya~line profile
    shapes of 91 emitters at $z = 3.1$. While we are unable to make a
    direct quantitative comparison between our asymmetry measurement
    with their symmetry index,~\cite{Yamada12} found that 27 of
    their objects are asymmetric (``strong red and weak blue''),
    though at least half are only moderately so. Only three of their
    objects have blue wings (``strong blue and weak red''), whereas
    the rest of their sample are slightly to moderately
    asymmetric. These results are qualitatively consistent with our
    observations that the majority of our sample are slightly
    asymmetric ($0 < s < 2$) with a tail trailing off to high skewness.

    The bottom panel in \fig \ref{fig:elongskew} illustrates the
    distribution in axial ratio for the LAEs. %, including the dropouts.   
    Again, the distributions appear
    similar across the subgroups. The median axial ratio decreases
    slightly from $B$-dropout ($\langle a/b \rangle = 1.22$) to
    $V$-dropout ($\langle a/b \rangle = 1.21$) and 
    $i$-dropout ($\langle a/b \rangle = 1.17$), but this may well be due to small number
    statistics since only 9 $i$-dropouts were identified in the
    sample. Indeed, the KS tests for the axial ratios measured for the
    different sub-groups have $p$-values of $0.60-0.83$, suggesting
    that within the limits of our tests, the distributions are not
    significantly different. 

    We further consider the observed axial ratio distributions
      of LBGs at these redshifts. ~\cite{Ferguson04} reported that the
    ellipticity distribution for a sample of $z \approx 4$ $B$-dropouts
    resembled that of flattened disks, reasoning that many of the LBGs
    have several concentrations of light spread across the galaxy
    instead of a nucleated concentration. The mean ratio they derived is
    $a/b = 1.54$, higher than that for our $B$-dropouts. The discrepancy
    between our measurements is likely due to the difference in sample
    selection. The~\cite{Ferguson04} sample of LBGs include LAEs as
    well as non-emitters, and the latter likely suffer from severe
    optical depth, absorption, or radiative transfer
    effects. Therefore, it is sensible that our LAEs would exhibit a
    lower typical axial ratio and appear more spheroid-like than an
    LBG sample that incorporates emitters and non-emitters alike. How
    this optical depth effect may play into the relation between axial
  ratio and morphology will be revisited in \S \ref{opticaldepth}.

    \subsection{Evolution with Redshift}
\begin{figure}[ht]
  \centering
  \includegraphics[width=0.46\textwidth]{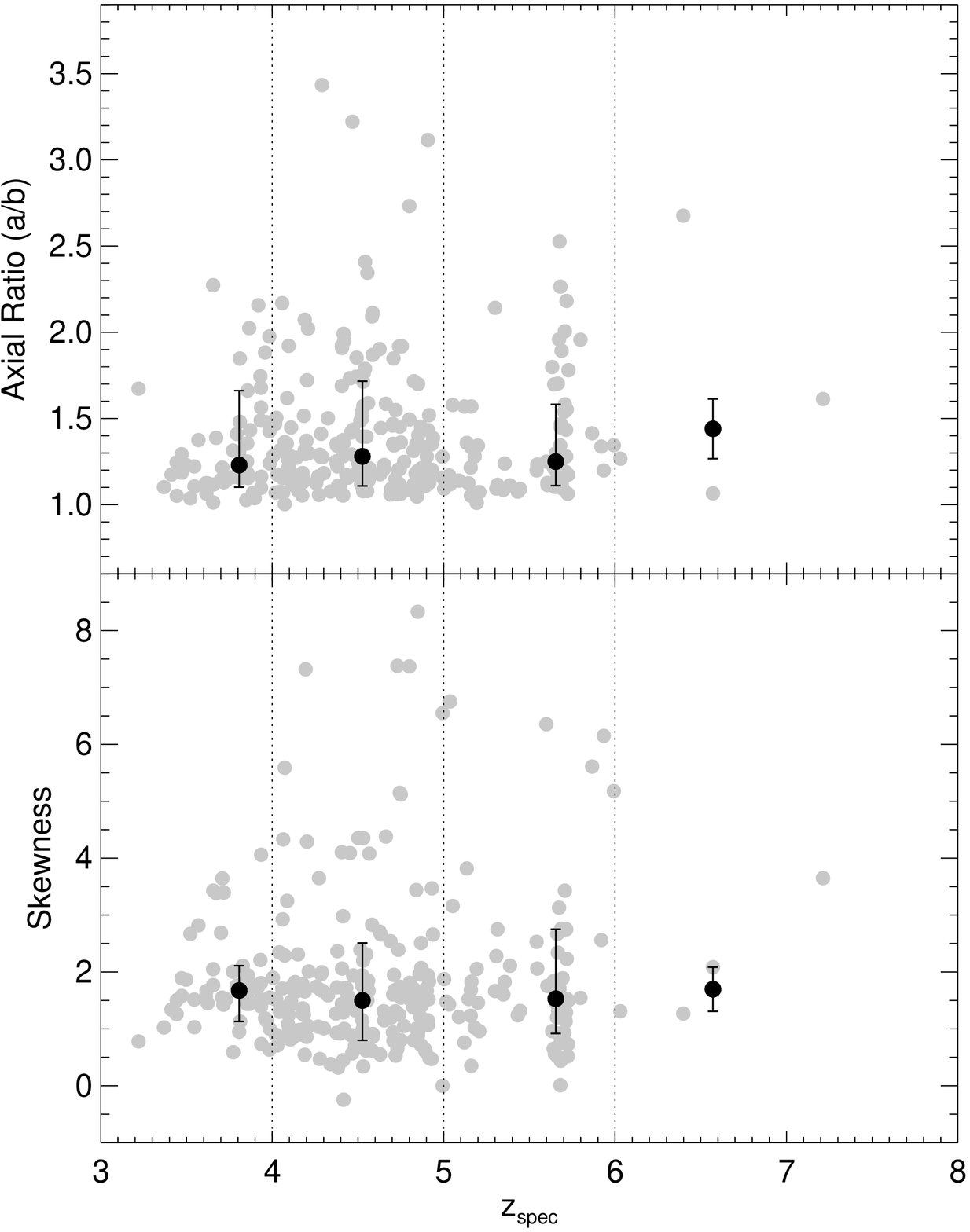}
  \caption{Axial ratio (top) and skewness (bottom) of the LAEs plotted
    as a function of redshift in different redshift bins (as
    segregated by the dotted lines). The grey
    points are from the full sample, overlaid by black filled circles
    that denote the median values of the axial ratio and skewness
    within each redshift bin. Vertical error bars
    illustrate the ranges where 68\% of the sources reside. No
    evolutionary trend is observed with either property, with Spearman
    rank coefficients $\rho_{a/b} = 0.05$ ($p$-value $=$ 0.41) and
    $\rho_{\rm Skew} = -0.004$ ($p$-value $=$ 0.94).}
  \label{fig:zelongskew}
\end{figure}

   \ref{tbl:lbgsummary} illustrates some differences in
    skewness and axial ratio among the $BVi$-dropouts, and this distinction
   raises the question of whether or not redshift-dependent factors
   such as IGM absorption 
    and diminishing sensitivity have an effect on the \lya~profile
    shape and morphology. In order to investigate if the trend of
    decreasing dispersion in skewness as a function of axial ratio is
    due more to intrinsic properties than redshift, we plot each
    of these properties as a function of redshift in
    \fig~\ref{fig:zelongskew}.  The full sample as well as median
    values of axial ratio and skewness within each redshift bin are
    plotted. We perform a Spearman rank correlation test as a
    nonparametric measure of potential statistical dependence between
    each of these properties with redshift, and determined $\rho_{a/b}
    = 0.05$ ($p$-value $=$ 0.41) and $\rho_{\rm Skew} = -0.004$
    ($p$-value $=$ 0.94). As such, we find no evolutionary trend with
    either property. Additional discussion on how IGM may affect the
    \lya~line profile can be found in \S \ref{IGM}. 

\begin{figure}[ht]
  \centering
  \includegraphics[width=0.45\textwidth]{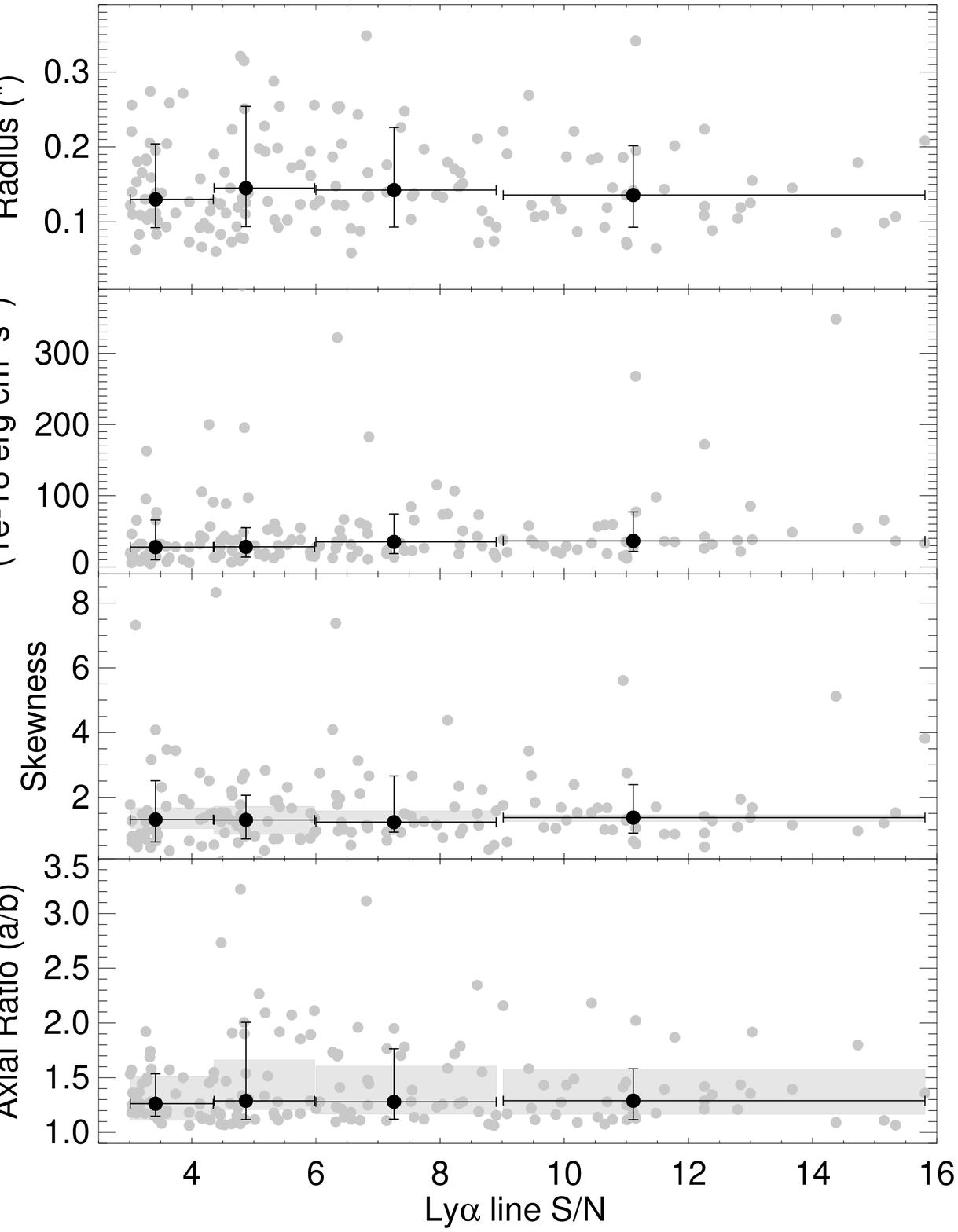}
  \caption{(Top to bottom) Source radius, \lya~flux,
    skewness, and axial ratio plotted as a function of S/N in
    \lya~line flux for the subsample of COSMOS LAEs with flux calibrations. The bins in SNR
    are equal in numbers, with their ranges represented by the horizontal
    error bars. Median values along with vertical error bars highlight the ranges where 68\% of the
    sources reside. In the bottom two panels,
      the grey rectangles represent the typical $\pm$ 1-$\sigma$
      uncertainties in skewness and axial ratio, respectively, on
      individual data points centered on the mean value of each
      bin. The error on axial ratios were not directly output from
      Source Extractor, but rather represent the propagated results on
      typical errors from Source Extractor measurements~\cite[][]{Ravindranath06}.}
  \label{fig:checksn}
\end{figure}

  Another check for the validity of the trends in axial ratio and
  skewness regards their correlation with the signal-to-noise ratio
  (SNR) cut applied to the initial selection of the LAE sample.  We
  focus on a subset of our LAEs in the COSMOS
  field~\cite[][]{Mallery12} with proper flux calibrations for this
  test. We observe a relatively flat distribution of source radii as a
  function of \lya~line SNR in
  \fig~\ref{fig:checksn}, with Spearman rank coefficient $\rho_{\rm
    radius} = -0.01$ and $p$-value $= 0.89$. We note an increase in \lya~line flux
  correlated with the sources with SNR $\gtrsim$ 6 ($\rho_{\rm flux} =
  0.34$ with $p$-value of $4.04 \times 10^{-6}$), but no strong trends can
  be seen in skewness ($\rho_{\rm Skew} = 0.13$ with $p$-value of
  0.08) or axial ratio ($\rho_{a/b} = -0.01$ with $p$-value of 0.92) with \lya~line SNR.  

  \section{Discussion}
  \label{Discussion}
  
  In this section we draw on the observed trends in skewness and axial
  ratio to infer the physical mechanism and geometric or projection
  effect for the propagation of
  \lya~photons. We start with the simplifying assumption that
  these star-forming galaxies are intrinsically round, symmetric
  disks. We note that while we treat the axial
  ratio as a proxy for the viewing angle of the galaxy, it would not
  be a perfect correlation without kinematics data to help disentangle
  the intrinsic inclination angle of the galactic disks. Another
  caveat is that although many of the sources with low axial ratio may
  simply be compact, the fraction of sources with low axial ratios
  (\eg $a/b < 1.5$) but which have half-light radii $r > 0\farcs15$ is
  nontrivial, $\sim 24\%$. Thus, we do consider the case where the
  axial ratio may reflect the viewing angle of the galaxy.
  In the following sections, we discuss two plausible
    scenarios for the origin of the \lya~photons, and explore how the
    line profile shape would vary in each case in the context of our
    observations. The overall impact from the IGM is also examined.

%  \subsection{Outflows and Distribution of \lya~emitting Clouds}
  \subsection{Outflows in High-Redshift LAEs}
  
  Outflows are known to be commonplace in high-redshift
  star-forming systems~\cite[][]{Pettini01,Frye02,Vanzella09}. In
  particular, many studies have presented observational spectroscopic
  signature consistent with outflows and expanding shells in $z \sim
  2-3$ LBG samples~\cite[\eg][]{Shapley03,Steidel10,Kulas12}.  It has been
  shown that the \lya~profile shape depends on the expanding velocity
  of the shell~\cite[][]{Verhamme06}, though this prediction is not currently
  tested for our sample until the systemic velocity is measured
  independently from the \lya~line. Here we consider
  the effect of interstellar and circumgalactic gas kinematics on the
  \lya~photons and galaxy viewing angle.

  If the \lya~emitting hydrogen clouds were behind outflowing
  gases perpendicular to the galactic disk, the broadening of the
  \lya~line together with intergalactic absorption could cause the
  line to appear asymmetric. Models from~\cite{Verhamme06} have shown
  that in the case of an expanding shell with a central monochromatic
  source, the more backscatterings the \lya~photons undergo on their
  escape path, the more the red wing will be broadened in the line 
  profile. This effect is most pronounced in the 
  face-on cases where we observe the outflowing gas along our line of
  sight, and is negligible if the outflow velocity vector is
  perpendicular to our line of sight. This is consistent with the
  trend seen in \fig \ref{fig:elongskew}, suggesting that the physical
  mechanism motivating the observed \lya~line profile in the context of the
  galaxy elongation may be outflows. At low
  redshifts,~\cite{Kornei12} found a correlation between inclination
  and outflow velocity for a sample of $z \sim 1$ star-forming
  galaxies, while~\cite{Law12} found none for their $z \sim 2$
  sample. An interesting follow-up question for the subset of high S/N
  sources in our sample is how the outflow velocities relate to
  \lya~profile and UV morphology.

  % The distribution of \lya~emitting regions, in particular with
  % respect to the dust patchiness in the galaxy, could play a role in
  % affecting the \lya~line profile as well.  All else being equal, the
  % \lya~line would appear single-peaked more often in the case of
  % homogeneously-distributed \lya~photons than centrally-located
  % ones~\cite[][]{Garavito14}. If the \lya~emitting gas is uniformly 
  % distributed across the galaxy, we may expect the line to appear more
  % skewed, particularly at low viewing angle in the optically thick
  % regime. On the contrary, if the sources radiating \lya~photons are
  % centrally concentrated, the line will more likely appear
  % double-peaked. The asymmetry thus measured will be lower as optical
  % depth increases, which is more consistent with the trend from our
  % observations.  In fact, using a nonparametric morphological
  % indicator such as the Gini coefficient,~\cite{Vanzella09} found that
  % LAEs at similar redshifts (especially
  % those with equivalent widths $>$ 20\AA) are intrinsicially
  % nucleated in the $z_{850}$ band, potentially suggesting that the
  % \lya~radiating sources are centrally concentrated. Thus, our
  % spectroscopic approach of investigating the profile shape offers an
  % independent way to probe the spatial distribution of the
  % \lya~emitting sources within the galaxies.

  \subsection{Rotation, Viewing Angle, and Optical Depth}
  \label{opticaldepth}
  We next consider the case where the \lya~emission emerges
    from the galactic disk, in the absence of outflows.
 In a recent paper, 
  \cite{Garavito14} presented the results of radiative transfer
  calculations measuring the impact of gas bulk rotation on
  the line profile shape. The fraction of \lya~photons that escape at
  various velocities (as a proxy for the \lya~line profile shape) is
  shown for different rotations and viewing angles. A double-peaked
  emission line may become single-peaked when the 
  photon-emitting source has large rotation velocity ($V = 200-300$ km
  s$^{-1}$). This effect is most pronounced for the edge-on view as
  well as low neutral hydrogen optical depth.
  An additional effect that may complicate the line morphology 
  is the absorption blueward of the \lya~peak by intergalactic gas
  clouds. (A more detailed discussion of IGM effects is in \S
  \ref{IGM}.) %Due to this absorption, the \lya~photons at ``negative''
%  velocities relative to the systemic velocity of the galaxy would not be seen. 
  Thus, a \lya~line that should
  otherwise have been double-peaked would now appear asymmetric,
  e.g. in the case of fast rotation ($V > 200$ km s$^{-1}$) at low
  viewing angle.  It becomes single-peaked in the case of minimal
  rotational velocity, regardless of the viewing angle.  This can be
  seen in both Figures 4 and 5 of~\cite{Garavito14}. 

  While these models may explain the relatively symmetric
  single-peaked lines for sources with high viewing angle and minimal
  rotation, they underpredict the asymmetric lines observed at low
  axial ratios, nor do they sufficiently explain the lack of
  asymmetric profile shapes at high axial ratios. To address these
  issues, we further consider the optical depths probed in these galaxies. The
  sources with low axial ratios are likely to exhibit lower optical depths
  as viewed face-on, contrary to those that appear edge-on. Looking at
  the models from~\cite{Garavito14}, if the viewing angle is held
  constant, say, in the face-on case (\eg $\lvert \cos(\theta) \rvert
  = 1$ in their \fig 2), the red part of the line would become more and more
  symmetric as the optical depth increases from $10^5$ to $10^7$. This
  may be due to the fact that, in the optically thick case, the
  \lya~photons are more likely to scatter into the wing of the line
  and appear double-peaked. In this light, the red part of the
  \lya~line would appear more symmetric at high $a/b$ values, leading to the
  relative lack of sources at the high axial ratio --- high skewness
  parameter space. 
  
  \subsection{Absorption by the Intergalactic Medium}
  \label{IGM}
  The effect of IGM absorption on the line profile of
  \lya~has been examined and modeled at length by previous
  studies. For instance,~\cite{Laursen11} computed the line
  profile shape as well as the escape fractions of \lya~photons for a
  suite of simulated galaxies with different circular
  velocities. While the \lya~line asymmetries was not explicitly
  quantified, the study investigated the relative strength of the two
  peaks in \lya~as viewed through different IGM conditions at various redshifts. The
  result that the blue peak is often weak, if not completely
  absorbed, at $z > 3.5$. This is consistent with our sample having very few
  double-peaked sources and mostly strong-red-peak sources.
  The~\cite{Laursen11} models show that some absorption on the blue
  side of the red peak may be due to the IGM, which could be
  responsible for the skewness we observed. However, many other
  factors (e.g. outflows) may be in play such that the profile shape,
  and hence \lya~asymmetry, would not necessarily increase with
  redshift, which is consistent with what we observed in \fig
  \ref{fig:zelongskew}. 

  \cite{Dijkstra07} also studied the impact of IGM on the profile
  shapes of high-redshift \lya. Using a skewness parameter originally
  defined by~\cite{Kashikawa06}, \cite{Dijkstra07} find that the
  observed shape of \lya~may vary widely depending on the star
  formation rate, ionizing background, and the intrinsic width and
  systemic velocity of the line. In particular, they find that 
  galaxies with large star formation rates or those embedded
    in a strong ionizing background tend to display symmetry in the \lya~line, at least
  near the low end of our redshift range. As our sample of GOODS
  and COSMOS LAEs were drawn from different areas in the sky, the
  effect of biased cosmic variance seems unlikely, as could be
  identified in our images. The models from~\cite{Dijkstra07} may
  imply that our LAEs have moderate star formation rates.

  \section{Conclusions and Summary}
  \label{Summary}
  The origin and escape path of \lya~photons in high-redshift galaxies
  has been a challenge to understand due to the complexity of the scattering and
  extinction process by the interstellar medium as well as absorption by 
  intergalactic gas and dust. An investigation of both the
  spectroscopic and morphological properties of a large sample of LAEs
  is necessary to understand the link between the morphology of the
  rest-frame UV emission and that 
  of the spectral profile shape of the \lya~emission line.
  In this paper we study the relationship between the asymmetry
  exhibited by the \lya~line profile and the rest-frame UV ($\sim
  2300$\AA) morphology of a spectroscopically selected sample of 304
  \lya~emitting sources at $3 < z < 7$ in the GOODS and COSMOS fields.  
  Within the sample, 66 $B$-dropouts, 43 $V$-dropouts, and 7 $i$-dropouts have
  been identified. The median redshift of the sample is $\langle z \rangle = 4.55$. 

  Our main results include the following:
  \begin{itemize}

    \item There is a large spread in skewness values
  for all sources at low axial ratio values, and this spread decreases
  as the elongation of the galaxies increases. This trend is
  consistent with the expectations that outflows are present. Models
  have shown that in the case of a galaxy with a central monochromatic
  source and an expanding shell of gas, \lya~photons would
  undergo backscatterings on their escape path and broaden the line
  profile, and this effect would be most pronounced in the face-on
  cases where the outflowing gas is directly along our line of sight. 
  If outflows were present, and indeed recent literature seems to
  suggest that they are commonplace in these high-redshift star-forming
  systems, the expected orientation of the galactic winds as well as
  the broadening effect on the spectral line are both consistent with
  the observed trend between \lya~asymmetry and UV morphology in this
  work.  %Additionally, whether the ionizing sources are centrally or
%  homogeneously distributed in a galaxy would affect the asymmetry of
%  the \lya~line at various viewing angles and optical depths. Our
%  spectroscopic approach to compare line profile shape with the theoretical
%  models suggests that the LAEs in our sample are likely to host
 % centrally-originated \lya~photons, in agreement with conclusions
 % drawn from independent morphological analyses in the literature.  

 \item In the absence of outflows, we explore the \lya~line
   profile shapes predicted by radiative
   transfer models that probe different rotation velocities and
   distributions of the ionizing sources at various optical depths and
   viewing angles. With the simplifying assumption that these
     LAEs are intrinsically round and symmetric disks, rotation
     alone is insufficient in explaining the observed trend between
     skewness and axial ratio. In order to account for the relative
     abundance of asymmetric line profiles in the low and high axial
     ratio cases, consideration of optical depths is also invoked.
     The optical depth probed may
  directly indicate the amount of intervening gas, and thus,
  absorption and scattering of \lya~photons into the line wings, along
  the line of sight. The enhancement of the double-peaked feature in
  the \lya~line would cause its profile shape to appear symmetric at
  high $a/b$ values.

  \item Both the \lya~line profile shape and the axial ratio of
    the rest-frame UV emission may potentially be redshift dependent
    due to IGM absorption and diminishing sensitivity,
    respectively. Results from our Spearman rank correlation tests
    show that neither axial ratio nor \lya~skewness has any
    statistical dependence on redshift, so any variations we see in
    each of the spectral and morphological properties are likely
    intrinsic to the galaxies. We also checked for any potential
    effect on source size, \lya~flux, skewness, and axial ratio that
    the SNR of the \lya~line may have, and found no correlation.
%    Our results, based on a combined spectroscopic and
%    morphological approach should be robust given that \lya~line is
%    adequately detected.

  % \item We examine the star-forming main sequence relation for our
  % LAEs. The relation is much weaker in the low redshift bins ($z < 4$)
  % than at higher redshifts. There appears to be negligible
  % differences when the sample is categorized by its axial ratio,
  % though sources with large skewness contribute to a larger scatter at
  % high redshifts.
 \end{itemize}

\acknowledgements

We thank the referee for a thorough review and many helpful
suggestions for improving the paper. VU acknowledges helpful
discussions of the various statistical tests with CWK Chiang, as well
as partial funding support from the Thirty Meter Telescope
International Observatory and the UC Chancellor's Postdoctoral Fellowship
Program. The  work of DS was carried out at Jet Propulsion Laboratory,
California Institute of Technology, under a contract with NASA. 
The data presented herein were obtained
at the W.M. Keck Observatory, which is operated as a scientific
partnership among the California Institute of Technology, the
University of California and the National Aeronautics and Space
Administration. The Observatory was made possible by the generous
financial support of the W.M. Keck Foundation.  
The authors wish to recognize and acknowledge the very significant
cultural role and reverence that the summit of Mauna Kea has always
had within the indigenous Hawaiian community.  We are most fortunate
to have the opportunity to conduct observations from this mountain. 

\bibliography{lae}
\bibliographystyle{apj}

\clearpage
%\LongTables
\begin{deluxetable*}{lccccccc}
   \centering
   \tabletypesize{\scriptsize}
%   \tabletypesize{\tiny}
   \tablewidth{0pt}
   \tablecolumns{8}
   \tablecaption{Spectroscopic Properties of LAEs}
   \tablehead{   % column headings
     \colhead{Source} &
     \colhead{RA (J2000)} &
     \colhead{Dec (J2000)} &
     \colhead{Type} &
     \colhead{Redshift} &
     \colhead{Redshift Flag} &
     \colhead{Skewness} &
     \colhead{Axial Ratio}
     	}
   \startdata
  1  &  53.14262 & -27.82654 & Bdrop & 3.569 & A & 2.82$\pm$0.10 & 1.38 \\ 
  2  &  53.23541 & -27.86569 & Other & 4.705 & B & 1.61$\pm$0.03 & 1.35 \\ 
  3  &  53.17774 & -27.82750 & Vdrop & 4.839 & A & 1.45$\pm$0.04 & 1.10 \\ 
  4  &  53.07359 & -27.89222 & Bdrop & 3.701 & A & 2.69$\pm$0.32 & 1.14 \\ 
  5  &  53.09012 & -27.95106 & Bdrop & 4.213 & A & 2.01$\pm$0.04 & 1.30 \\ 
  6  &  53.08168 & -27.81108 & Bdrop & 3.715 & A & 1.43$\pm$0.06 & 1.15 \\ 
  7  &  53.09518 & -27.74385 & Bdrop & 4.694 & A & 1.20$\pm$0.12 & 1.09 \\ 
  8  &  53.06141 & -27.79965 & Other & 5.934 & A & 6.15$\pm$1.53 & 1.20 \\ 
  9  &  53.06190 & -27.78507 & Other & 4.531 & A & 4.35$\pm$0.35 & 1.12 \\ 
 10  &  53.08402 & -27.82344 & Bdrop & 3.655 & A & 2.05$\pm$0.20 & 2.27 
\enddata
   \label{tbl:specprop}
   \tablenotetext{1}{Table \ref{tbl:specprop} is published in its
     entirety in the electronic edition of ApJ. A portion is shown
     here for guidance regarding its format and content.}
   \tablenotetext{2}{For the GOODS fields, quality flag `A' indicates
     the most reliable redshifts, whereas the `B' redshifts are most
     likely correct. See text for details. Redshifts for the COSMOS sources have previously been published in
     \cite{Mallery12}.}
 \end{deluxetable*}

\begin{deluxetable*}{lcccccc}
   \centering
   \tabletypesize{\scriptsize}
%   \tabletypesize{\tiny}
   \tablewidth{0pt}
   \tablecolumns{7}
   \tablecaption{Summary of $BVi$ Dropouts}
   \tablehead{   % column headings
     \colhead{Type} &
     \colhead{\# in GOODS} &
     \colhead{\# in COSMOS} &
     \colhead{\# Total} &
     \colhead{$\langle z_{\rm spec} \rangle$} &
     \colhead{$\langle$Skew$\rangle$}  & 
     \colhead{$\langle a/b \rangle$}
     	}
   \startdata
All & 130 & 174 & 304 & 4.55 & 1.55 & 1.26 \\
$B$-dropout & 57 & 9 & 66 & 3.93 & 1.58 & 1.21 \\
$V$-dropout & 20 & 23 & 43 & 4.82 & 1.68 & 1.22 \\
$i$-dropout & 6 & 1 & 7 & 5.61 & 1.81 & 1.25
%Other & 60 & 141 & 201 & 
\enddata
   \label{tbl:lbgsummary}
 \end{deluxetable*}

\clearpage

\appendix

\section{Testing the Robustness of our Line-Fitting Program}
\begin{figure}[ht]
  \centering
  \includegraphics[width=0.5\textwidth,angle=90]{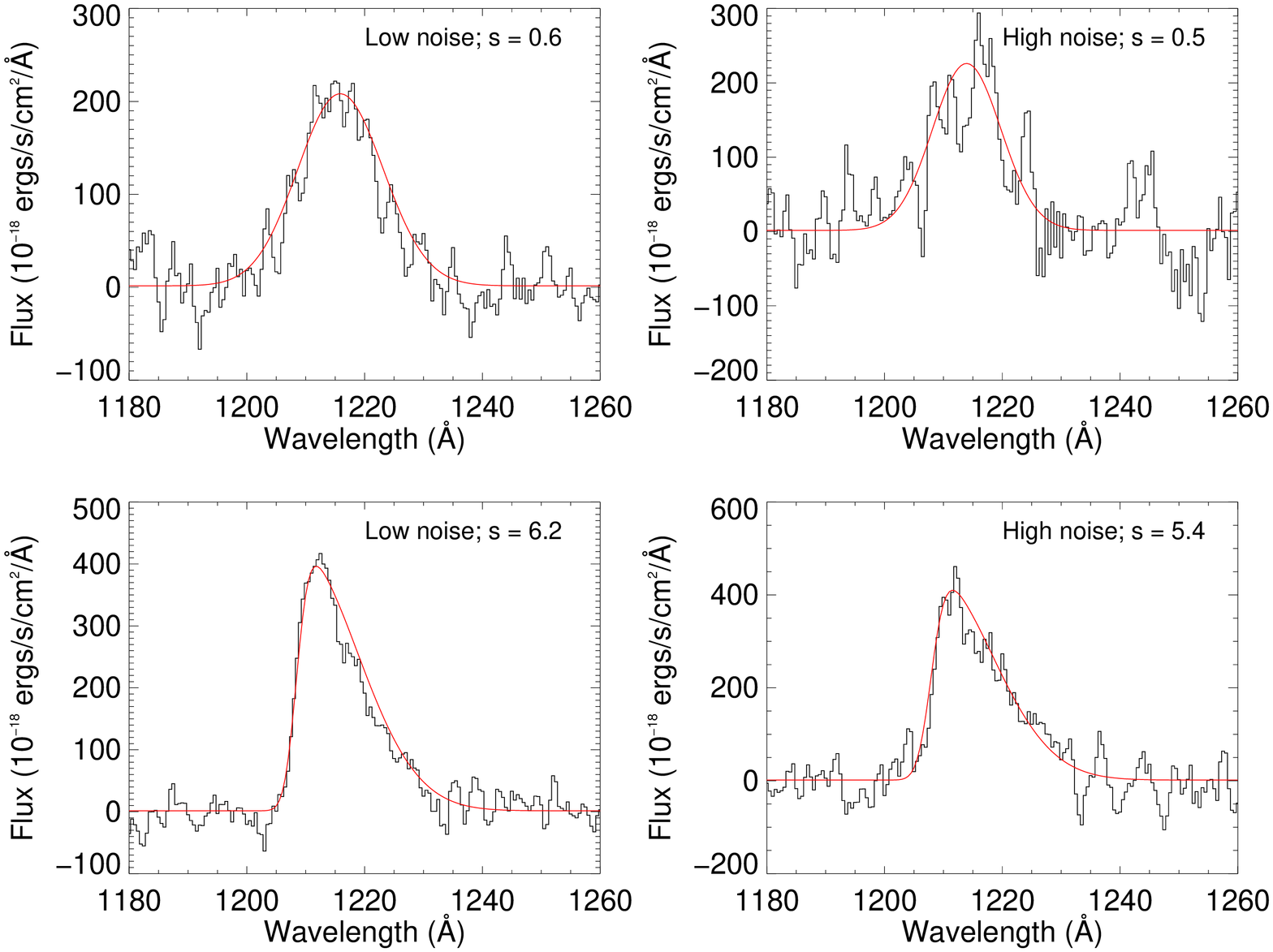}
  \caption{Examples of simulated spectra (black) with low and high
    intrinsic skewness and noise levels, respectively. The panels in
    the top row have intrinsic skewness of $s = 0.5$, and those in the
    bottom row have $s = 6.5$. The spectra have been smoothed by a
    width of 3 pixels for presentation purposes. The skewed Gaussian fits are shown in red.} 
  \label{fig:simlines}
\end{figure}

To test the robustness of our skewness measurements from our
line-fitting procedure, particularly at the low S/N level, we show
here an in-depth simulation exercise. Using the Gaussian equation
given in Eqn. \ref{eqn:gauss}, we simulated a suite of Gaussian lines
with varying skewness ($s = 0.5, 2.5, 4.5, {\rm and}~6.5$) at two
different noise levels (see examples in \fig \ref{fig:simlines}). Random noise has been added to the simulated
spectra with fixed amplitude, centroid, line width, and continuum
level. This is done so that the effect of noise on skewness
measurement can be isolated. For our final set of simulations, 10
spectra have been generated at each of four skewness and two noise
levels, totaling a suite of 80 simulated spectra to be tested with our
line-fitting program.  

\begin{figure}[ht]
  \centering
  \includegraphics[width=0.4\textwidth,angle=90]{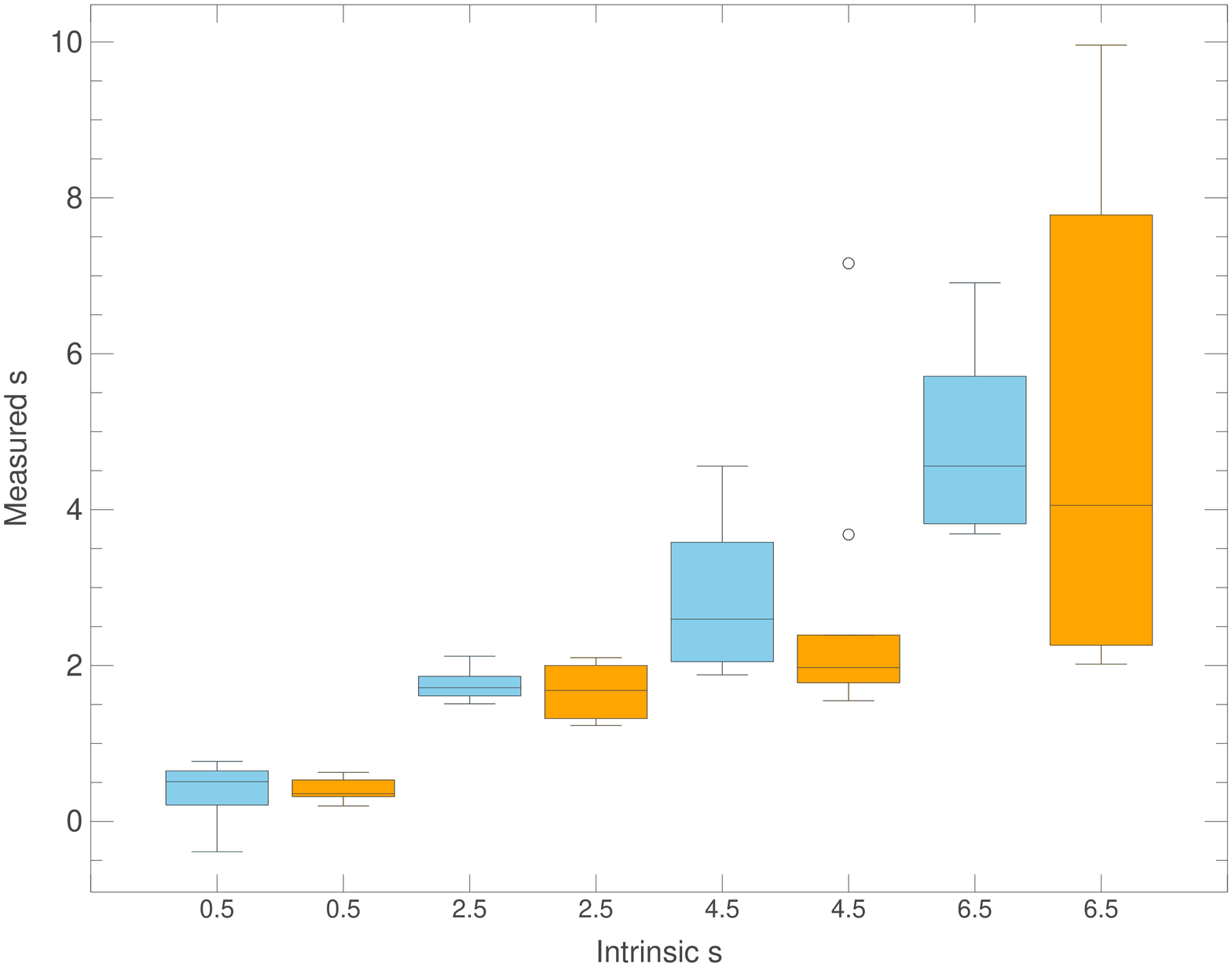}
  \caption{A box and whisker plot showing the interquartile (25$^{th}$ and
    75$^{th}$) and max/min ranges of the measured skewness of the simulated
    spectra in low (pale blue) and high (orange) noise bins. Over all,
    it appears that the measured skewness falls short of the
    corresponding intrinsic skewness. At low skewness values, 
    noise level plays a less significant role in affecting the
    resulting fits. Lines that are more skewed tend to be fitted with
    a lower skewness value as noise increases. A couple
    outlier points are identified with small circles.} 
  \label{fig:boxplot}
\end{figure}

Our main fitting results from the simulation exercise is summarized in
\tbl \ref{tbl:sim} and \fig \ref{fig:boxplot}. In general, the
measured skewness values fall short of the corresponding intrinsic
parameter across the span of the skewnesses probed. For low skewness
values, noise level plays a less significant role in affecting the
resulting fits. Lines that are intrinsically more skewed tend to have
their skewness underestimated as noise increases, though the
differences in the median and mean skewnesses between low and high
noise levels are marginal (up to 8\% at $s = 6.5$). 

These results imply that our skewness values as measured
from our line-fitting program may be underestimated, depending on the
noise level. Given the lack of trends seen in skewness versus
\lya~line S/N for our sample (\fig \ref{fig:checksn}), we believe that any systematics that
may be inherent in our fitting procedure would not affect our
overall conclusions and in particular, \fig \ref{fig:elongskew}, in
the paper. In fact, this exercise offers confidence that our highly
asymmetric data points are indeed real, preserving the discussions of
the major results in this work. 

\begin{deluxetable*}{lcccc}[hb]
   \centering
   \tabletypesize{\scriptsize}
%   \tabletypesize{\tiny}
   \tablewidth{0pt}
   \tablecolumns{5}
   \tablecaption{Skewness Measurements of Simulated Spectra}
   \tablehead{   % column headings
     \colhead{Noise Level} &
     \colhead{$s = 0.5$} &
     \colhead{$s = 2.5$} &
     \colhead{$s = 4.5$} &
     \colhead{$s = 6.5$} 
     	}
   \startdata
Low: 10\% of Amp. & $\langle s \rangle = 0.51$ & $\langle s \rangle = 1.72$ & $\langle s \rangle = 2.60$ & $\langle s \rangle = 4.56$ \\
                 & $\overline{s} = 0.40 \pm 0.33$ & $\overline{s} = 1.75 \pm 0.18$ & $\overline{s} = 2.87 \pm 0.90$ & $\overline{s} = 4.85 \pm 1.07$ \\
High: 20\% of Amp. & $\langle s \rangle = 0.36$ & $\langle s \rangle = 1.68$ & $\langle s \rangle = 1.98$ & $\langle s \rangle = 4.06$ \\
                 & $\overline{s} = 0.41 \pm 0.13$ & $\overline{s} = 1.66 \pm 0.32$ & $\overline{s} = 2.65 \pm 1.61$ & $\overline{s} = 4.81 \pm 2.74$ 
\enddata
   \label{tbl:sim}
 \end{deluxetable*}

% \begin{deluxetable*}{ccc}
%   \centering
%   \tabletypesize{\scriptsize}
%   \tablewidth{0pt}
%   \tablecolumns{3}
%   \tablecaption{Median Masses and Star Formation Rates (SFRs)}
%   \tablehead{
%     \colhead{} &
%     \colhead{log(Mass / $M_\odot$)} & 
%     \colhead{log(SFR / [$M_\odot$ yr$^{-1}$])} 
%   }
%   \startdata
%    $a/b < \langle a/b \rangle $ & 9.57$\pm$0.63 & 1.27$\pm$0.69 \\
%    $a/b > \langle a/b \rangle $ & 9.71$\pm$0.55 & 1.26$\pm$0.65 \\
%    Skewness $<$ $\langle$Skewness$\rangle$ & 9.80$\pm$0.54 & 1.41$\pm$0.62 \\ 
%    Skewness $>$ $\langle$Skewness$\rangle$ & 9.45$\pm$0.61 & 1.23$\pm$0.70 
%   \enddata
%   \label{tbl:ms}
% \end{deluxetable*}

\end{document}